\documentclass[onecolumn,showpacs,superscriptaddress,nofootinbib,10pt]{revtex4-1}

\usepackage{graphicx}
\usepackage{amssymb}
\usepackage{amsmath,float,upgreek,mathtools}
\usepackage[mathscr]{euscript}

\def\rU{\textrm{r}_{\textrm U}}
\def\rH{\textrm{r}_{\textrm H}}
\def\rh{\textrm{r}_{\textrm h}}
\def\be{\begin{equation}}
\def\ee{\end{equation}}

\newcommand{\beq}{\begin{equation}}
\newcommand{\eeq}{\end{equation}}

\newcommand{\ba}{\begin{eqnarray}}
\newcommand{\ea}{\end{eqnarray}}

 \def\de{\partial} \def\lb{\rangle}
\def\>{\rangle} 
\def\<{\langle} 
   
\newcommand\orcidroldao{{\href{https://orcid.org/0000-0003-3978-532X}{\orcidicon}}}
\newcommand{\orcidicon}{%
	\begin{tikzpicture}
	\draw[lime, fill=lime] (0,0)
		circle [radius=0.16]
		node[white] {{\fontfamily{qag}\selectfont \tiny ID}};
	\draw[white, fill=white] (-0.0625,0.095)
		circle [radius=0.007];
	\end{tikzpicture}	\hspace{-2mm}
}

\usepackage{color,graphicx}
\usepackage[colorlinks,hyperindex]{hyperref}
\definecolor{green1}{RGB}{0,128,0} 
\hypersetup{
  colorlinks = true,
  linkcolor  = blue,
  citecolor = cyan,
}
\usepackage{bookmark,textgreek}
\usepackage{hyperref,color,xcolor}
\usepackage{tensor,tikz}
\definecolor{green1}{RGB}{0,128,0} 
\hypersetup{hidelinks,colorlinks=true,breaklinks=true,urlcolor= blue}

\usepackage{tensor}

\begin{document}
 
\title{MGD-decoupled black holes, anisotropic fluids and holographic entanglement entropy}

\author{Rold\~ao da Rocha\orcidroldao\!\!}
\email{roldao.rocha@ufabc.edu.br}\affiliation{Center of Mathematics, Federal University of ABC, Santo Andr\'e, Brazil}
\author{Anderson A. Tomaz}
\email{anderson\_tomaz@id.uff.br}\email{anderson.tomaz@ufabc.edu.br}\affiliation{Center for Natural and Human Sciences, Federal University of ABC, Santo Andr\'e, Brazil}\affiliation{Institute of Physics, Fluminense Federal University, Niter\'oi, Brazil}


\begin{abstract}
The holographic entanglement entropy (HEE) is investigated for a black hole under the minimal geometric deformation (MGD) procedure, created by gravitational decoupling via an anisotropic fluid, in an AdS/CFT on the brane setup. The respective HEE corrections are computed and confronted to the corresponding corrections for both the standard MGD black holes and the Schwarzschild ones. 
\end{abstract}

\maketitle

\section{Introduction}
\label{intro}

The method of geometric deformation (MGD)  consists of a protocol to derive compact stellar configurations of the effective Einstein's field equations  on the brane \cite{Casadio:2015gea,Ovalle:2017wqi,Ovalle:2017fgl,covalle2,Ovalle:2014uwa,Ovalle:2016pwp,GCGR,CoimbraAraujo:2005es,Ovalle:2017wqi,Ovalle:2017fgl,covalle2}. The MGD is a well succeeded theory that 
allows the study of nonlinear gravity in braneworlds, whose effective action can be obtained at low energies. There is a precise and intrinsic relationship between Gauss-Codazzi-like geometrical methods  and AdS/CFT, as comprehensively paved in Refs. \cite{Kanno:2002iaa,Soda:2010si}. This approach also includes dark radiation, that naturally arises as homogeneous solutions. In this setup, the bulk gravity is dual to CFT on the brane, providing  a holographic interpretation of braneworld scenarios as underlying apparatuses to MGD \cite{Kanno:2002iaa,Soda:2010si}.

 The MGD, and its extensions \cite{Casadio:2012pu,Casadio:2012rf,Casadio:2015gea,Ovalle:2013vna}, comprise high precision phenomenological bounds that physically regulate their inherent parameters. The strictest bounds on the brane tension were derived in Ref. \cite{Casadio:2015jva,Casadio:2016aum,Fernandes-Silva:2019fez}. In addition, hydrodynamical analog systems, that emulate  MGD black holes in the laboratory, were studied in Ref. \cite{daRocha:2017lqj}. Besides, MGD black strings were proposed in Ref. \cite{Casadio:2013uma}. Refs. \cite{Contreras:2018gzd,Ovalle:2007bn,Singh:2019ktp,Sharif:2019mzv,Casadio:2019usg,Rincon:2019jal,Hensh:2019rtb,Ovalle:2019lbs,Gabbanelli:2019txr,Contreras:2019mhf,Ovalle:2019qyi,Ovalle:2018ans,Sharif:2018tiz,Ovalle:2018vmg,Morales:2018urp,PerezGraterol:2018eut,Morales:2018nmq,Contreras:2019iwm,Contreras:2019fbk}  include and study anisotropic solutions of quasi-Einstein's equations, in the context of the MGD procedure  \cite{Fernandes-Silva:2017nec}. Besides, anisotropic MGD-decoupled solutions were  obtained by gravitational decoupling methods \cite{Gabbanelli:2018bhs,Panotopoulos:2018law,Heras:2018cpz,Contreras:2018vph,Ovalle:2017khx,Ovalle:2013xla}. The MGD was also studied in the context of the strong gravitational lensing \cite{Cavalcanti:2016mbe}, whereas MGD glueball stars were scrutinized in Refs. \cite{daRocha:2017cxu,Fernandes-Silva:2018abr}. In addition,  MGD black holes in the GUP context  were studied in Ref. \cite{Casadio:2017sze}, and relativistic anisotropic compact stellar configurations have been recently derived in \cite{Tello-Ortiz:2020euy}.  

The MGD-decoupling method was later introduced when one  iteratively produces, from a source of gravity, more intricate, weakly-coupled,  gravitational sources, that still preserve spherical symmetry \cite{Ovalle:2017fgl}. Once the MGD decoupling is introduced by a perfect fluid via the brane effective Einstein's equations, additional sources that are weakly-coupled to gravity   induce anisotropy. When a perfect fluid couples to more intricate stress-energy tensors that describe matter-energy in more realistic setups, interesting phenomena appear. Compact stellar configurations do not necessarily request the  isotropic condition -- when the radial pressure, $p_r$, equals the tangential pressure, $p_t$. Indeed, when matter  that constitute the stellar configuration is denser  than nuclear matter, anisotropic equations of state (EoS) set in. Therefore, anisotropy in compact stellar configurations takes into account more  realistic scenarios. Refs. \cite{Ovalle:2017fgl,Ovalle:2019lbs,Tello-Ortiz:2019gcl,Ovalle:2019qyi,Ovalle:2017wqi} shows that for positive anisotropy, when $ p_t - p_r > 0$, the compact stellar configuration repulsive force   counterbalances the gravitational pressure. Hence, anisotropic stars are more likely to exist in astrophysics, being also more stable, as discussed in Refs.  \cite{Ivanov:2018xbu,Sharif:2018toc,Stelea:2018cgm,Heras:2018cpz,Ovalle:2018umz,Ivanov:2018xbu,Estrada:2018zbh,Jasim:2018wtd,Maurya:2019hds,Cedeno:2019qkf}.  A formidable step in this direction is the recent direct observation of anisotropic compact configurations, through the detection of gravitational waves. The Tolman-Oppenheimer-Volkoff equations can represent balance  conditions for  compact stellar configurations. However, the EoS is necessary for describing the complete structure of these compact stellar configurations. Experimental data shows the existence of such astrophysical objects, observed at very high densities, including X-ray pulsars, bursters and sources \cite{Maurya:2019noq,Maurya:2019wsk}.
Recently, strange stars candidates, illustrated by the astrophysical SAX J1808.4-3658  compact stellar configurations, were described by the anisotropic MGD-decoupling \cite{Tello-Ortiz:2019gcl}. In addition, anisotropic neutron compact stellar configurations were used to describe the compact objects 4U 1820.30, 4U 1728.34, PSR J0348+0432, RX J185635.3754, PSR 0943+10, the binary pulsar SAX J1808.4-3658 and X-ray binaries Her X-1 and Cen X-3, whose stability was also investigated in the MGD-decoupling context \cite{Torres:2019mee,Deb:2018ccw}. A similar procedure was scrutinized to  describe anisotropic color-flavor strange quark stars \cite{Lopes:2019psm}.

A relevant paradigm comprising the entanglement between states, yields a (nonlocal) correlation among quantum subsystems that are far apart from each other. A measure of quantum entanglement, known as entanglement entropy (EE), has been a spot of applications in quantum information, condensed matter, general relativity (GR), high energy physics. The most influential development in this field consists of the holographic entanglement entropy,  put forward by Ryu and
Takayanagi (RT) \cite{Ryu:2006bv}. It plays an important role on inspecting several facets of quantum entanglement, in strongly coupled QFTs, which 
represents dual theories to (weakly-coupled) gravity in codimension one bulk. The EE a any quantum system is utilized for restoring and reconstructing the geometrical constituent of a bulk. The holographic entanglement entropy (HEE) has been used for computing the EE of subsystems whose habitat is the dual theory. Since the celebrated RY expression takes into account minimal surfaces, analyzing them in several asymptotically AdS spacetimes is a relevant tool 
\cite{Hubeny:2007xt,Emparan:2006ni}. 

Based on Ref. \cite{Li:2010dr}, which defines the HEE in asymptotically flat spacetimes, the aim of this work consists to implement the procedure established in \cite{Sun:2016dch,daRocha:2019pla} to compute the HEE for a spacetime obtained by the MGD gravitational decoupling protocol, using an additional anisotropic fluid in the stress-energy tensor  \cite{Ovalle:2018umz}. Into this incursion, one intends to verify how two ways of positioning the boundaries -- either far from or almost on the horizon -- affects the HEE, up to the second-order in an expanded solution. This can answer if the first law of black hole thermodynamics still holds, as well as how to fit each order of the HEE correction when compared with HEE corrections arising from the standard MGD spacetime and the Schwarzschild one  \cite{daRocha:2019pla}.

This paper is organized as follows: in Sect.~\ref{sect:MGD-setup} the MGD-decoupling is applied to anisotropic black holes. The HEE for static, spherically symmetric, spacetimes is then discussed, via the RT formula. The computations of the HEE corrections, for MGD-decoupled spacetimes generated by anisotropic fluids, are derived and discussed in Sect.~\ref{sect:HEEtoMGD-anis} in two important regimes,  considering the boundary far from the event horizon and almost on it. More analysis and discussions about these results are scrutinized in Sect.~\ref{sect:Final}.

\section{MGD-decoupling and anisotropic black holes}
\label{sect:MGD-setup}
The MGD procedure can be realized as a mechanism that is usually employed to derive high energy corrections to GR. Denoting by $\sigma$ the brane tension, systems with energy $E\ll\sigma$ neither feel the self-gravity effects nor the bulk effects, which  then  allows the recovery of GR in such energies. An infinitely rigid brane scenario, representing the 4-dimensional GR  brane pseudo-Riemannian manifold, can be  implemented in the  $\sigma\to\infty$ limit. 
The most strict brane tension bound, $\sigma \gtrsim  2.81\times10^{-6} \;{\rm GeV^4}$, was derived in the extended MGD-decoupling context \cite{Fernandes-Silva:2019fez}. 

Starting from any straightforward static, spherically symmetric, source of gravity, corresponding to some stress-energy tensor ${\rm T}_{\mu\nu}$, one can iteratively introduce more intricate gravitational sources. This procedure is described by \cite{Ovalle:2017fgl,Ovalle:2017wqi} 
\begin{eqnarray}
\label{dt0}
{\rm T}_{\mu\nu}\mapsto \mathring{\rm T}_{\mu\nu}={\rm T}_{\mu\nu}+\sum_{m=1}^\infty\upalpha^{(m)}\,{\rm T}^{(m)}_{\mu\nu}
\ ,
\end{eqnarray}
where $\upalpha^{(m)}$  are constant parameters that drive the weakly-coupled effects of the sources ${\rm T}^{(m)}_{\mu\nu}$. 
This procedure holds whenever the backreaction among the sources is disregarded, namely $
\nabla^{\mu} T^{(m)}_{\mu\nu}=0$, for any natural number $m$. Hereon natural units will be used.

The MGD decoupling for a perfect fluid can be introduced by considering  Einstein's equations
\begin{equation}
\label{corr2}
R_{\mu\nu}-\frac{1}{2}\,R\, g_{\mu\nu}
=
-\upkappa^2\,\mathring{\rm T}_{\mu\nu},
\end{equation}
where $\upkappa^2$ denotes the Newton coupling constant. One assumes that the total stress-energy tensor is endowed with two contributions, 
\begin{equation}
\label{emt}
\mathring{\rm T}_{\mu\nu}
=
{\rm T}_{\mu\nu}+\upalpha\,\Uptheta_{\mu\nu}
\ ,
\end{equation}
where $
{\rm T}_{\mu \nu }=(\uprho +p)\,u_{\mu }\,u_{\nu }-p\,g_{\mu \nu }$ 
is the 4-dimensional  stress-energy tensor for a perfect fluid with 4-velocity field $u^\mu$,
density $\uprho$ and pressure $p$.
The term $\Uptheta_{\mu\nu}$ in Eq.~(\ref{emt}) describes an additional source whose weakly-coupling to gravity is driven by $\upalpha$, inducing anisotropy. 
As the Einstein tensor does satisfy the Bianchi identity, the source in Eq.~(\ref{emt}) also satisfies the Eq. (\ref{dt0}). 

Given a static, spherically symmetric, metric  
\begin{equation}
ds^{2}
=
-e^{\upnu (r)}\,dt^{2}+e^{\uplambda (r)}\,dr^{2}
+r^{2}d\theta^{2}+r^2\sin ^{2}\theta \,d\phi ^{2}
\ ,
\label{metric}
\end{equation}
the fluid 4-velocity reads $u^{\mu }=e^{-\upnu /2}\,\delta _{0}^{\mu }$, 
for $r\in[0,R]$, where $R=\int_0^\infty \rho(r') r'^{3}\,dr'/\int_0^\infty \rho(r') r'^{2}\,dr'$ emulates the compact star surface radius.
The metric~(\ref{metric}) must satisfy Einstein's equations~(\ref{corr2}),
yielding
\begin{eqnarray}
\label{ec1}
\upkappa^2
\left(
\uprho+\upalpha\,\uptheta^0_{\ 0}
\right)
&=&
\frac 1{r^2}
+
e^{-\uplambda }\left(\frac{\uplambda'}r- \frac1{r^2}\right)\ ,
\\
\label{ec2}
\upkappa^2
\left(-p+\upalpha\,\uptheta^1_{\ 1}\right)
&=&
\frac 1{r^2}
-
e^{-\uplambda }\left(\frac{\upnu'}r+ \frac 1{r^2}\right)\ ,
\\
\label{ec3}
4\upkappa^2
\left(p-\upalpha\,\uptheta^2_{\ 2}\right)
&=&
e^{-\uplambda }
\left(2\,\upnu''-\upnu'^2-\uplambda'\,\upnu'
+\frac{2(\upnu'-\uplambda')}r\right)
\ ,
\end{eqnarray}
where $f'= df/dr$. Besides, the equality $\uptheta^3_{\ 3}=\uptheta^2_{\ 2}$ comes from spherical symmetry.
Eq. (\ref{dt0}) yields \cite{Ovalle:2017fgl,Ovalle:2017wqi} 
\begin{equation}
\label{con1}
p'
+
\frac{\upnu'}{2}\left(\uprho+p\right)
+\upalpha\left[
\frac{\upnu'}{2}\left(\uptheta^0_{\ 0}-\uptheta^1_{\ 1}\right)
+\left(\uptheta^1_{\ 1}\right)'+
\frac{2}{r}\left(\uptheta^2_{\ 2}-\uptheta^1_{\ 1}\right)\right]
=
0
\ ,
\end{equation}
The limit $\upalpha\to 0$, corresponding to the perfect fluid case, is then formally redeemed.
\par
One can define the effective density, and the effective radial and tangential pressures, respectively, by   \cite{Ovalle:2017fgl,Ovalle:2017wqi} 
\begin{eqnarray}
\mathring{\uprho}=
\uprho
+\upalpha\,\uptheta^0_{\ 0}, \qquad \qquad 
\mathring{p}_{r}
=
p-\upalpha\,\uptheta^1_{\ 1}, \qquad\qquad  \mathring{p}_{t}
=
p-\upalpha\,\uptheta^2_{\ 2}.
\end{eqnarray}
It leads the $\Uptheta_{\mu\nu}$ tensor to induce a coefficient of anisotropy of a compact stellar configuration, 
\begin{equation}
\label{anisotropy}
\Updelta
\equiv
\mathring{p}_{t}-\mathring{p}_{r}
=
\upalpha\left(\uptheta^1_{\ 1}-\uptheta^2_{\ 2}\right).
\end{equation}
\par
The MGD-decoupling can now be applied to the case at hand by simply noting that the
stress-energy tensor~\eqref{emt} is precisely of the form~\eqref{dt0}, with
${\rm T}_{\mu\nu}$ as the one of a perfect fluid, $\upalpha^{(1)}=\upalpha$ and ${\rm T}_{\mu\nu}^{(1)}=\Uptheta_{\mu\nu}$, being ${\rm T}_{\mu\nu}^{(m)}\equiv0$, for all $m\geq 2$.
The components of the diagonal metric $g_{\mu\nu}$ that solve the complete Einstein
equations~\eqref{corr2} and satisfy the MGD read~
$
\mathring{g}_{\mu\nu}
=
{g}_{\mu\nu}={g}_{\mu\nu}^{(1)}
$, for {\small{$\mu=\nu\neq 1$}}, and
$
\mathring{g}^{11}
=
{g}^{11}+\upalpha\,g^{(1)11}$ \cite{Ovalle:2017fgl,Ovalle:2017wqi}. Hence, solely the radial metric  component carries signatures of $\Uptheta_{\mu\nu}$. One can solve the Einstein's equations for a perfect fluid ${\rm T}_{\mu\nu}$,
\begin{equation}
\label{f1}
{G}_{\mu\nu}
=
-\upkappa^2\,{\rm T}_{\mu\nu}
\ ,
\qquad
\qquad
\nabla^\mu {\rm T}_{\mu\nu}=0,
\end{equation}
and then the remaining quasi-Einstein equations for the source $\Uptheta_{\mu\nu}$,
\begin{equation}
\mathring{G}_{\mu\nu}
=
-\upkappa^2\,{\Uptheta}_{\mu\nu}
\ ,
\qquad\qquad
\nabla^\mu\Uptheta_{\mu\nu}=0,
\label{MGD2}
\end{equation}
where the divergence-free quasi-Einstein tensor  $
\mathring{G}_{\mu\nu}
=
{G}_{\mu\nu}+\Upgamma_{\mu\nu}$, 
with $\Upgamma_{\mu\nu}=\Upgamma_{\mu\nu}(g_{\rho\sigma})$ denotes a metric dependent tensor that is   
 divergence free \cite{Ovalle:2017wqi,Ovalle:2018umz}. 
\par
A feasible solution of the coupled system in \eqref{f1}, for a perfect fluid, 
reads 
\begin{equation}
ds^{2}
=
-e^{\upxi (r)}\,dt^{2}
+
\frac{dr^{2}}{\upmu(r)}
+
r^{2}d\theta^{2}+r^{2}\sin ^{2}\theta \,d\phi ^{2}
\ ,
\label{pfmetric}
\end{equation}
where 
\begin{equation}
\label{standardGR}
\upmu(r)
=
1-\frac{\upkappa^2}{r}\int_0^r \tilde{r}^2\,\uprho(\tilde{r})\, d\tilde{r}
=
1-\frac{2M(r)}{r}
\end{equation}
$m(r)$ denotes the Misner-Sharp mass, $r$-dependent, function, which by  measuring the amount of energy within a sphere of areal radius $r$, provides a coherent quasilocal definition to the curvature-producing energy in  black holes. 
The source $\Uptheta_{\mu\nu}$ effects on the perfect fluid solution $\{\upxi,\upmu,\uprho,p\}$ can be then encoded in the MGD into the radial component of the perfect fluid geometry in (\ref{pfmetric}).
Namely, the general solution is given by Eq.~\eqref{metric} with $\upnu(r)=\upxi(r)$ and
\begin{eqnarray}
\label{expectg}
e^{-\uplambda(r)}
=
\upmu(r)+\upalpha\,{\rm f}^\star(r)
\ ,
\end{eqnarray}
where ${\rm f}^\star={\rm f}^\star(r)$ is the MGD function to be determined from Eqs.~\eqref{MGD2}, given by \cite{Ovalle:2018umz}
\begin{eqnarray}
\label{ec1d}
\upkappa^2\,\uptheta^0_{\ 0}
&=&
-\frac{{\rm f}^\star}{r^2}
-\frac{{\rm f}^{\star'}}{r}
\ ,
\\
\label{ec2d}
\upkappa^2\,\uptheta^1_{\ 1}
&=&
-{\rm f}^\star\left(\frac{1}{r^2}+\frac{\upxi'}{r}\right)
\ ,
\\
\label{ec3-dimensional }
4\upkappa^2\,\uptheta^2_{\ 2}
&=&
-{{\rm f}^\star}\left(2\,\upxi''+\upxi'^2+\frac{2\upxi'}{r}\right)
-{{\rm f}^{\star'}}\left(\upxi'+\frac{2}{r}\right)
\ .
\end{eqnarray}

Black hole solutions are derived from an EoS when one determines the vacuum MGD function ${\rm f}^\star$.
The MGD metric will therefore read
\begin{equation}
\label{Schw}
ds^2
=
-\left(1-\frac{2M}{r}\right)dt^2
+\frac{dr^2}{\strut\displaystyle{1-\frac{2M}{r}+\upalpha\,{\rm f}^\star(r)}}
+r^2(\mathrm{d}\theta^2+\sin^2\theta \mathrm{d}\varphi^2).
\end{equation}
\par
For the Schwarzschild solution, the surface $\rH=2M$
is both a Killing horizon and an exterior marginally trapped
surface \cite{Ovalle:2018umz}.
For the MGD-decoupled Schwarzschild metric~\eqref{Schw}, the component $g_{tt}(r)=e^{\upnu(r)}$ 
equals the Schwarzschild standard form, having coordinate singularity at  $r=\rH$, being also a Killing horizon. 
However, the causal horizon at $r=\rh$ is such that $g^{rr}(\rh)=e^{-\uplambda(\rh)}=0$, or equivalently, 
\begin{equation}
\label{newBH}
\rh
\left[1+\upalpha\,{\rm f}^\star(\rh)\right]
=
2M
\ .
\end{equation}
One then demands that $\rh\ge 2M$, in such a way that the surface $r=\rH$ is either concealed behind, or at most coincides with, the causal horizon. Eq. \eqref{Schw}, corresponding to the MGD-decoupled metric, can represent  a black hole only if the causal and the Killing horizons coincide, namely, $\rh=2M=\rH$. This condition will be assumed in what follows.

Now let one considers an anisotropic configuration, governed by a EoS of type  \cite{Ovalle:2018umz} 
\begin{equation}
\label{generic}
\Uptheta_0^{\,0}
=
a\,\Uptheta_1^{\,1}+b\,\Uptheta_2^{\,2}
\ ,
\end{equation}
with $a$ and $b$ constants.
Conformal configurations are obtained by setting $a=-1$ and $b=-2$, whereas barotropic configurations require the choice $a=-1/K$ and $b=0$ \cite{Ovalle:2018umz}.
Eqs.~(\ref{ec1d} -- \ref{ec3-dimensional }) then yield the ODE for the MGD function
\begin{eqnarray}
\label{giso}
&&{\rm f}^{\star'}\left[\frac{1}{r}-\frac{b}{4}\left(\upxi'+\frac{2}{r}\right)\right]
+ {\rm f}^\star\left[\frac{1}{r^2}-a\left(\frac{1}{r^2}+\frac{\upxi'}{r}\right)
-\frac{b}{4}\left(2\,\upxi''+\upxi'^2+\frac{2\upxi'}{r}\right)\right]
=
0,
\end{eqnarray}
whose general solution for $r>\rH=2M$ is given by
\begin{equation}
{\rm f}^\star(r)
=
\left(1-\frac{2M}{r}\right)
\left(\frac{\ell}{r-BM}\right)^{\!\!A}
\ ,
\label{giso2}
\end{equation}
where $\ell>0$ represents a length scale, and
\begin{eqnarray}
\label{AB}
A
&=&
\frac{2\,(a-1)}{b-2}>0, \qquad\quad
B
=
\frac{b-4}{b-2}
\ ,
\end{eqnarray}
with obviously $b\neq 2$.  The range $A>0$ is required for deriving an asymptotic flat solution. Hence, 
\begin{equation}
\label{Gsol}
e^{-\uplambda(r)}
=
\left(1-\frac{2M}{r}\right)
\left[1+\upalpha
\left(\frac{\ell}{r-BM}\right)^{\!\!A}\right].
\end{equation}
\par
The effective
density, and the radial and tangential pressures are respectively given by 
\begin{eqnarray}
\mathring{\uprho}
&=&
\upalpha\,\uptheta^0_{\ 0}
=
-\frac{\upalpha}{\upkappa^2\,r^2}
\left(\frac{\ell}{r-BM}\right)^{\!\!A}
\left[1-A\left(\frac{2M-r}{BM-r}\right)\right],
\label{Gefecden}\\
\mathring{p}_{r}
&=&
-\upalpha\,\uptheta^1_{\ 1}
=
\frac{\upalpha}{\upkappa^2\,r^2}
\left(\frac{\ell}{r-BM}\right)^{\!\!A},
\label{Gefecprera}\\
\mathring{p}_{t}
&=&
-\upalpha\,\uptheta^2_{\ 2}
=
-\frac{\upalpha\,A}{2\,\upkappa^2\,r^2\,\ell}\left(r-M\right)
\left(\frac{\ell}{r-BM}\right)^{\!\!A+1}.
\label{Gefecptan}
\end{eqnarray}
They diverge at $
r_{\rm c}
=
BM,$ that is a physical singularity at the range $0<r_{\rm c}<\rH$ for $0<B<2$. For $B>2$, $r_{\rm c}$ represents a physical singularity outer  the Killing horizon, $\rH$, being this case  physically forbidden, to preclude naked singularities.
Moreover, the effective radial and tangential pressures are related by
\begin{equation}
\label{rel}
\mathring{p}_t
=
-\frac{A}{2}\left(\frac{r-M}{r-BM}\right)
\mathring{p}_r
\ .
\end{equation}
Since $A>0$, one concludes that both the radial and tangential pressures do  have contrary signs. In addition, the MGD effective density and radial pressure satisfy 
\begin{equation}
\mathring{\uprho}
=
\left[A\left(\frac{r-2M}{r-BM}\right)-1\right]
\mathring{p}_r\label{rel3}\sim
\left\{
\begin{array}{ll}
-\mathring{p}_r
&
{\rm for}\
r\sim\,2M
\\
\left(A-1\right)\,\mathring{p}_r
&
{\rm for}\ 
r\gg 2M
\ .
\end{array}
\right.
\end{equation}
Since $A>0$, the asymptotic behaviour in Eq.~\eqref{rel3} demands the range $A\in(0,1]$, for the density does not change sign in the sector $r\in(2M,+\infty)$. 
For $\upalpha$ negative, the effective density is positive.
\par
For $b\in(2,4)$, the coordinate singularity $r_{\rm c}$ attains  negative values. Hence, no extra singularity, besides the usual Schwarzschild one at $r=0$, exist, if there are no solutions, $r_0>0$, of the equation $e^{-\uplambda(r)}=0$ \cite{Ovalle:2017wqi}.
This is satisfied whenever $\upalpha>0$, for any $A>0$, that is, for $a>1$.
There is also a second coordinate singularity, solution of $e^{-\uplambda(r)}=0$, consisting of 
\begin{equation}
\label{Gsz}
r_0
=
BM+{\ell}\,(-\upalpha)^{1/A}
>
r_{\rm c}
\ ,
\end{equation}
when $\upalpha<0$.
To produce  a physically viable black hole solution, this solution $r_0$ have to  attain lower values than the existing singularity.
If $b\in(2,4)$, then the existing singularity $r=0$ yields  
$
r_0
\le
0,
$. Equivalently, if $\ell$ and $|\upalpha|$  satisfy
\begin{equation}
{\ell}\,(-\upalpha)^{1/A}
\le
-BM
\ .
\label{clinL}
\end{equation}
Otherwise, if $b<0$ or $b>4$, the relevant singularity occurs at $0<r_{\rm c}<\rH$, but 
$r_0>r_{\rm c}$ makes this case inappropriate. 
The final conclusion is thus that the linear EoS~\eqref{generic} always produces black holes, with a Schwarzschild-like physical singularity at $r=0$,  if $b\in(2,4)$ and $a>1$.

Not being so generalist, and to further explore the physical 
content of this model, one adopts hereon $a=2$ and $b=3$, implying that 
$A=2$ and $B=-1$. This choice allows no other additional physical singularity than  the well known  $r=0$ one. In fact, Refs. \cite{Ovalle:2017fgl,Ovalle:2017wqi} analyzed the particular case $b = 3$ with $a > 1$, to ensure that solutions are 
asymptotically  flat.

\section{HEE for a black hole from anisotropic fluid under MGD}\label{sect:HEEtoMGD-anis}

The entanglement entropy (EE), ${\rm S}_A$, of some manifold $A$, with boundary $\de A$,  
is an important quantity in 4-dimensional  QFTs. Indeed, it represents the von Neumann entropy of the (reduced) density matrix, in the case where degrees of freedom, into a 3-dimensional  space-like submanifold, $B$, are stretched out \cite{Ryu:2006bv}. 
The EE quantifies how $A$ is correlated to $B$, measuring the amount of  entropy in $A$, by an observer isolated from 
$B$. Therefore, there is a part of the  
AdS$_5$ bulk from which one can compute ${\rm S}_A$ in the gauge/gravity duality. At a zero temperature regime, any system in QFT is characterized by a pure state $|\Psi\lb$, where the respective density
matrix reads $ \rho=|\Psi\rangle \langle
\Psi|.$  When one splits the quantum system into $A$ and $B$, an  observer isolated from $B$ describes the quantum system by a reduced density matrix $\rho_A= \mathrm{Tr}_{B}~\rho.$ 
Defining the EE of $A$ as 
${\rm S}_A =
- \mathrm{Tr}_{A}\,
\rho_{A} \log \rho_{A}$, if the density matrix $\rho$ is pure, then
$ S_A=S_B$.   This equality is violated by finite temperature QFTs. 
The inequality $
 S_{A+B}\le S_{A}+S_{B}$ is always valid. Considering a QFT on a 4-dimensional  spacetime splitting, 
$\mathbb{R}\times \Sigma_3$, into some timelike direction and a 3-dimensional 
 spacelike manifold, $\Sigma_3$, a 3-dimensional submanifold $B\subset
\Sigma_3$ represents a geometric complement of 
$A$. An ultraviolet cut-off, ${\rm b}$, prevents the EE to diverge.
The coefficient that drives this divergence at the continuum limit depends on  the area of $\de A$, 
\be {\rm S}_A\propto \frac{\mbox{area}(\de
A)}{{\rm b}^{2}}.
 \label{divarea}\ee
The
conformal Poincar\'e metric, that characterizes  AdS$_5$ bulk geometry, reads 
\be ds^2=\frac{R^2}{z^2}\left(dz^2+dx_\mu dx^\mu \right). 
\label{Poincare} \ee
The CFT$_{4}$ that is dual to gravity in the bulk lives on the  $\mathbb{R}^{1,3}$ boundary, at $z=0$, with coordinates $(x^0,x^i)$. The AdS$_{5}$ conformal  coordinate, $z$, is the energy scale.

Besides the pure AdS$_5$ spacetime (\ref{Poincare}), AdS$_5$ black holes can be  also regarded. 
Indeed, the boundary $\partial A$ of $A$ can be extended to some manifold $\gamma_A$, whose boundary equals $\partial A$. Consequently, the EE ${\rm S}_A$ in the  CFT$_{4}$ reads \cite{Ryu:2006bv,Hubeny:2007xt,Emparan:2006ni}. 
\begin{equation} {\rm S}_{A}=\frac{{\rm area}(\gamma_{A})}{4\upkappa^2_{5}},
\label{eq:HEE}
\end{equation}
where $\upkappa^2_{5}$ is the 5-dimensional  Newton coupling constant. 
\begin{figure}[H]
    \centering
    \includegraphics[scale=0.55]{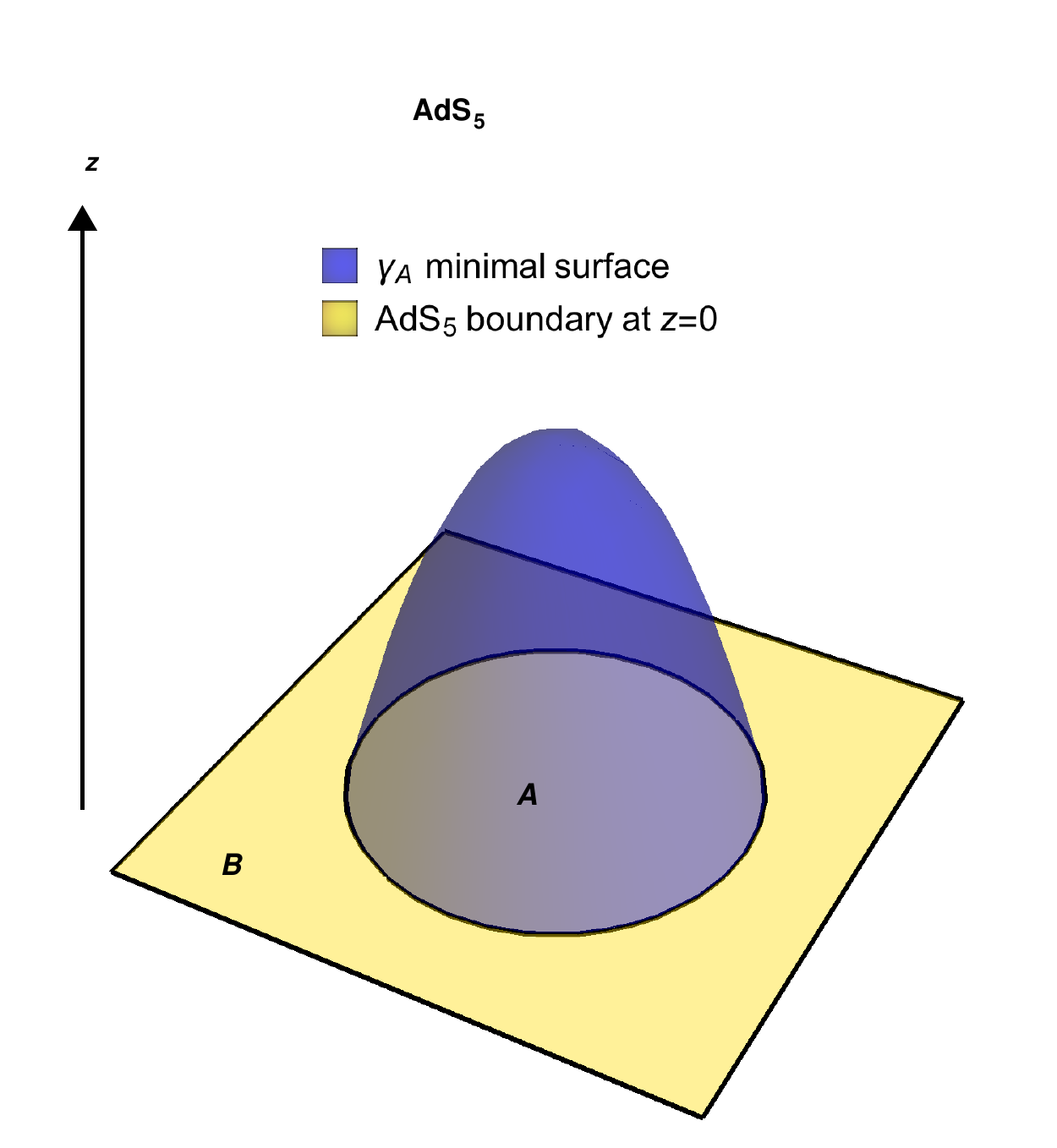}
    \caption{The blue minimal surface $\gamma_A$, where codim($\gamma_A) = 2$ with respect to the AdS$_5$ bulk, is moored at the boundary $\partial A$ of the (entangling) sector $A$, whose area yields its EE.
}
    \label{fig:fig}
\end{figure}

\subsection{HEE corrections}

\subsubsection{Far from the horizon}\label{subsect:MGDanisFar}
For computing  HEE corrections, the first procedure consists to settle the boundary manifold in a sector that is very far away from the event horizon, named $\textrm{r}=\textrm{r}_{\textsc{L}}$, which is still at a finite large distance. Now, let a circle in polar coordinates, be defined by the angle $\uptheta=\uptheta_0$, carrying the role to confine the entangling surface. Additionally, let $\textrm{r}=\textrm{r}(\uptheta)$ be the areal radius coordinate describing the minimal surface, whose boundary depicts the the entanglement manifold.

One needs to derive the minimum of the area function, 
\begin{equation}\label{eq:area}
\mbox{Area}(\gamma_A)={2\pi}\int_0^{\uptheta_0}\mathrm{d}\uptheta\,\textrm{r}\sin\uptheta\sqrt{
\textrm{e}^{\uplambda(\textrm{r})}\left(\frac{\mathrm{d}\textrm{r}}{\mathrm{d}\uptheta}\right)^{2}+\textrm{r}^2},
\end{equation}
where the boundary condition, $\textrm{r}(\uptheta_0)=\textrm{r}_{\textsc{L}}$, which is required  to find the minimal surface.

 Eq.~\eqref{eq:area} can be rewritten, when one substitutes  $y=\cos\uptheta$,  reading
\begin{equation}\label{eq:area2}
\mbox{Area}(\gamma_A)=\int_{y_0}^1 \mathrm{d}y \, \mathcal{I}= {2\pi}\int_{y_0}^1 \mathrm{d}y\textrm{r}\sqrt{\textrm{r}^2+(1-y^2)\mathcal{G}\dot{\textrm{r}}^2},~
\end{equation}
where $y_0=\cos\uptheta_0$, the derivative with respect to $y$ was denoted by a dot and $\mathcal{G}=\mathcal{G}\left(\textrm{r}(y)\right)\equiv\exp\left[{\uplambda\left(\textrm{r}(y)\right)}\right]$. After finding the global minimum of Eq.~\eqref{eq:area2}, one obtains the HEE by directly employing Eq.~\eqref{eq:HEE}.

Varying Eq.~\eqref{eq:area2} with respect to $\textrm{r}(y)$ yields 
\begin{equation}\label{eq:EOM}
(y^2-1)\left[2\mathcal{G}\textrm{r}^2\ddot{\textrm{r}}-2y\mathcal{G}^2\dot{\textrm{r}}^3+\left(\textrm{r}\frac{\mathrm{d}\mathcal{G}}{\mathrm{d}\textrm{r}}-6\mathcal{G}\right)\textrm{r}\dot{\textrm{r}}^2\right]
+4y\mathcal{G}\textrm{r}^2\dot{\textrm{r}}+4\textrm{r}^3=0~,
\end{equation}
 which is a ordinary differential equation (ODE) that is highly nonlinear, whose  most simple solution is derived by attributing $\mathcal{G}\equiv\mathcal{G}\left(\textrm{r}(y)\right)=1$, \emph{i.e.}, $\textrm{r}(y)= k_0/y$. 
According to \cite{Sun:2016dch}, it is possible to attain nontrivial solutions of  Eq.~\eqref{eq:EOM} through series expansions for both the $\mathcal{G}\left(\textrm{r}(y)\right)$ and $\textrm{r}(y)$ functions, respectively expressed by
\begin{subequations}
\begin{eqnarray}
\mathcal{G}\left(\textrm{r}(y)\right)&=&1-\sum_{n=1}^{\infty} \textrm{g}_n(y)\upepsilon^n~,\label{eq:Gexp}\\
\textrm{r}(y)&=&\frac{k_0}{y}+\sum_{n=1}^{\infty}\textrm{r}_n(y)\upepsilon^n~,\label{eq:rexp}
\end{eqnarray}
\end{subequations}
considering $\upepsilon$ as a dimensionless small parameter encoding  black hole mass $M$-to-$\textrm{r}_\textsc{L}$ ratio, that is, $\upepsilon = \frac{M}{\textrm{r}_\textsc{L}}.$ Just the $\mathcal{O}(\upepsilon)$ terms in \eqref{eq:Gexp} carry corrections when one considers a collapsing black hole case.

To determine the $\mathcal{G}$ function, up to the $2^{\rm nd}$-order, one has the following metric of this specific spacetime permeated by anisotropic fluid discussed in Sect. \ref{sect:MGD-setup}:
\begin{equation}\label{metricAnisEspec}
    \mathrm{d}s^2 = -\left(1-\frac{2M}{\textrm{r}}\right)\mathrm{d}t^2
+\frac{\mathrm{d}\textrm{r}^2}{\left(1-\frac{2M}{\textrm{r}}\right)
\left[1-\omega^2
\left(\frac{1}{\textrm{r}+M}\right)^{2}\right]}+\textrm{r}^2\mathrm{d}\Upomega^2~,
\end{equation}
where $\mathrm{d}\Upomega^2=\mathrm{d}\uptheta^2+\sin^2\uptheta \mathrm{d}\varphi^2$~ and one denotes $\upalpha\ell^2=-\omega^2$.
Therefore, one can achieve the $\textrm{g}$-functions, displayed in Eq.~\eqref{eq:Gexp}, \emph{i.e.},
\begin{subequations}\label{g-functions}
\begin{eqnarray}
\textrm{g}_1(y)&=&-\frac{2y\textrm{r}_\textsc{L}}{k_0}~,\\
\textrm{g}_2(y)&=&-\frac{y^2 \textrm{r}_\textsc{L}}{k_0^2}\left[\textrm{r}_\textsc{L}(4+\psi^2)-2\textrm{r}_1(y)\right]~,
\end{eqnarray}
\end{subequations}
It is important to stress out that a dimensional analysis was necessary, to indicate the MGD parameter linked to the expansion parameter $\upepsilon$ as $\omega=\psi M$. Clearly, it is always possible to construct higher orders  terms in Eq.~\eqref{eq:Gexp}, whenever necessary. Just for the current goal, one stopped at the second order. As it was demonstrated in \cite{Sun:2016dch,daRocha:2019pla}, Eq. \eqref{eq:EOM} is solved -- order by order -- employing $\textrm{g}_1(y)$ and $\textrm{g}_2(y)$ in Eq. \eqref{g-functions}.

To determine the modifications to the HEE up to the $2^{\rm nd}$-order, one needs to compute the $\textrm{r}$-functions as follows. Henceforth, one gets $1^{\rm st}$-order terms in $\upepsilon$ after expanding Eq.~\eqref{eq:EOM}, and using the $\textrm{g}$-functions listed in Eq.~\eqref{g-functions}. Hence,  the $1^{\rm st}$-order ODE can be expressed as
\begin{equation}\label{r1EDOaniso}
\ddot{\textrm{r}}_1(y) + \frac{\left(5 y^2-3\right)}{y \left(y^2-1\right)}\dot{\textrm{r}}_1(y)+\frac{\left(3 y^2-1\right)}{y^2\left(y^2-1\right)}\textrm{r}_1(y)-\frac{\left(3 y^2+1\right)\textrm{r}_\textsc{L}}{y^2 \left(y^2-1\right)}=0~.
\end{equation}
The resolution of Eq.~\eqref{r1EDOaniso} brings the constant of integrations $\textsc{A}_1$ and $\textsc{A}_2$, which have their values determined by precluding any type of singularity at $y=1$, \emph{i.e}., $y=\cos\uptheta\in\left[\cos\uptheta_0,1\right]$. Thus, it is imperative to establish $\textsc{A}_2=2\textrm{r}_\textsc{L}$. Next, the boundary condition $\textrm{r}_1(y)=0$ implies that $\textsc{A}_1=-\left(y_0+2\log[y_0/(1+y_0)]\right)\textrm{r}_\textsc{L}$. So, the first $\textrm{r}$-function reads
\begin{equation}\label{r1aniso}
    \textrm{r}_1(y)=\frac{\textrm{r}_\textsc{L}}{2}\left[y-y_0-2\log\left(\frac{1+y}{1+y_0}\right) +2\log\left(\frac{y}{y_0}\right)\right]~.
\end{equation}
Importantly, there is a subtle restriction due to limitations in the perturbative expansion, as aforementioned in Ref. \cite{Sun:2016dch}. In this sense, one enforces that, once $y=0$ cannot be ever achieved, then $\textrm{r}_1(y)$ is well defined in the range $\uptheta_0<\pi/2$ or, in an equivalent way, $y\in (0,1)$.

Now, one goes to the $2^{\rm nd}$-order in $\upepsilon$, proceeding as early and employing the $\textrm{r}_1(y)$ function, displayed in Eq. \eqref{r1aniso}, to determine
\begin{equation}\label{r2EDOaniso}
    \ddot{\textrm{r}}_2(y)+\frac{\left(5 y^2-3\right)}{y\left(y^2-1\right)} \dot{\textrm{r}}_2(y)+\frac{\left(3 y^2-1\right)}{y^2 \left(y^2-1\right)}\textrm{r}_2(y) - \frac{2\textrm{r}_\textsc{L}^2}{k_0}\left[\frac{(\psi^2 -1)y^3+3y-4}{y^2 \left(y^2-1\right)}\right]=0~.
\end{equation}
Proceeding analogously as it has been made to obtain $\textrm{r}_1(y)$, one has  
\begin{equation}\label{r2aniso}
    \textrm{r}_2(y)=\frac{\textrm{r}_\textsc{L}^2}{4yk_0}\left[(\psi^2+1)(y^2-y_0^2)+2(\psi^2-9)\log\left(\frac{y}{y_0}\right)+32\log\left(\frac{1+y}{1+y_0}\right)\right]~.
\end{equation}
 Again, the computation of the constant of integration, which is important to reach the $\textrm{r}$-function above, was realized by eliminating the divergences at $y=1$ and the boundary condition $\textrm{r}_2(y_0)=0$. 

Finally, one realizes the expansion $\mathcal{I}=\mathcal{I}_0+\upepsilon\mathcal{I}_1+\upepsilon^2\mathcal{I}_2$, within the area formula showed in Eq. \eqref{eq:area2}. Henceforth, the $\textrm{r}$-functions Eq.~(\ref{r1aniso}, \ref{r2aniso}) need to be engaged to calculate each order of contribution for the HEE, \emph{i.e.}, $\mathcal{S}^{\scalebox{.5}{\textrm{Anis}}}=\mathcal{S}_0^{\scalebox{.5}{\textrm{Anis}}}+\mathcal{S}_1^{\scalebox{.5}{\textrm{Anis}}}+\mathcal{S}_2^{\scalebox{.5}{\textrm{Anis}}}+\cdots$, where we stand ourselves up to the $2^{\textrm{nd}}$-order term.

The expression for the $0^{\textrm{th}}$-order is written as
\begin{equation}\label{S0anisoBeyond}
    \mathcal{S}_0^{\scalebox{.5}{\textrm{Anis}}}=\frac{A_0}{4}=\frac{1}{4}\int_{y_0}^0dy\mathcal{I}_0=\int_{y_0}^0dy \frac{2 \pi  k_0^2}{y^3}=\frac{\pi  k_0^2}{4}\left(-1+\frac{1}{y_0^2}\right)~.
    \end{equation}
Besides, the expression for the $1^{\textrm{st}}$-order reads
\begin{eqnarray}\label{S1anisoBeyond}
    \mathcal{S}_1^{\scalebox{.5}{\textrm{Anis}}}&=&\frac{A_1}{4}=\frac{\upepsilon}{4}\int_{y_0}^0dy\mathcal{I}_1=\frac{\pi \textrm{r}_\textsc{L}M}{2}\left(y_0-1\right)^2~,
\end{eqnarray}
whereas the $2^{\textrm{nd}}$-order is given by 
\begin{eqnarray}\label{S2AnisBeyond}
    \mathcal{S}_2^{\scalebox{.5}{Anis}}&=&\frac{A_2}{4}\nonumber\\&=&\frac{\upepsilon^2}{4}\int_{y_0}^0dy\mathcal{I}_2=\frac{\pi}{8} \left\{-\omega^2\left(1-y_0^2+2\log y_0\right)+M^2\left[(1-y_0)(y_0-7)+2\log y_0+16\log\left(\frac{2}{1+y_0}\right)\right]\right\}~.\nonumber\\
\end{eqnarray}


Emulating some of the useful results in Refs. \cite{Sun:2016dch,daRocha:2019pla}, one notices no difference confronting the $0^{\textrm{th}}$-order terms of HEE for the MGD, the Schwarzschild or the spacetime permeated by an anisotropic fluid, thus $\Xi_0=\mathring{\Xi}_0=1$. In addition, one recalls that
\begin{align}
    \mathcal{S}_1^{\scalebox{.5}{\textrm{MGD}}}&=(2-\upxi)\frac{\pi\textrm{r}_\textsc{L}M}{4}\left(y_0-1\right)^2~,\nonumber\\
    \mathcal{S}_2^{\scalebox{.5}{\textrm{MGD}}}&=\frac{\pi M^2}{32} \left\{ \left[2\upxi(13\!-\!3 y_0)\!-\!(\upxi^2\!+\!4)(7\!-\!y_0)\right](1-y_0) +16(\upxi-2)^2\log \left(\frac{2}{1+y_0}\right)+\left[(\upxi-2)^2-2\upxi\right]\log(y_0)\right]~,\nonumber\\
    \mathcal{S}_1^{\scalebox{.5}{\textrm{Schw}}}&=\frac{\pi \textrm{r}_\textsc{L}M}{2}\left(y_0-1\right)^2~,\nonumber\\
    \mathcal{S}_2^{\scalebox{.5}{\textrm{Schw}}}&=\frac{\pi M^2}{8} \left[(7-y_0)(y_0-1)+16\log \left(\frac{2}{1+y_0}\right)+2\log y_0\right]~,
\end{align}
where $\upxi$ is a parameter characterizing a pure MGD spacetime which carries the brane tension signature \cite{daRocha:2019pla,daRocha:2012pt,Abdalla:2009pg}.

 A comparative analysis can be established for the HEE corrections of those three spacetimes mentioned in the last paragraph. Henceforth, one defines the ratio, when likening the $n^{\rm th}$-order corrections of HEE for the spacetime described by an anisotropic fluid under the MGD approach and the pure MGD one, 
\begin{equation}\label{Xi_n}
    \Xi_n=\frac{\mathcal{S}_ n^{\scalebox{.5}{\textrm{Anis}}}}{\mathcal{S}_n^{\scalebox{.5}{\textrm{MGD}}}}~,
\end{equation}
where $\mathcal{S}_ n^{\scalebox{.5}{\textrm{Anis}}}$ and $\mathcal{S}_n^{\scalebox{.5}{\textrm{MGD}}}$ are the $n^{\rm th}$-order corrections of HEE for those spacetimes early mentioned. On the other hand, when $\upxi=0$, then $\mathcal{S}_n^{\scalebox{.6}{\textrm{MGD}}}=\mathcal{S}_n^{\scalebox{.5}{\textrm{Schw}}}$, which is the $n^{\rm th}$-order correction of HEE for the Schwarzschild spacetime. For this specific case, one defines  $\mathring{\Xi}_n=\mathcal{S}_ n^{\scalebox{.5}{\textrm{Anis}}}/\mathcal{S}_n^{\scalebox{.5}{\textrm{Schw}}}$~.


The next step can be accomplished, through  the calculation of the quantities displayed in Eq.~\eqref{Xi_n}. Thereupon, in a straightforward way, one gets $\Xi_0=1$,~$\mathring{\Xi}_0=1$ and
\begin{equation}\label{Xi_1}
    \Xi_1=\frac{1}{1-\frac{\upxi}{2}}~,
\end{equation}
while $\mathring{\Xi}_1=1$. Since $\upxi<0$, a higher value of the brane tension induces a lower $\Xi_1$~.

To continue the comparative analysis with the MGD spacetime and, consequently, with Schwarzschild ones, let one resets $\omega=\psi M$~ to determine
\begin{equation}\label{Xi_2}
    \Xi_2=\frac{4\left(y_0-1\right)\left[7+\psi^2+\left(\psi^2-1\right)y_0\right]-2\left(\psi^2-1\right)\log y_0 +16\log\left(\frac{2}{y_0+1}\right)}{\left(1-y_0\right)\left[-28+26\upxi-7\upxi^2+\left(4-6\upxi+\upxi^2\right)y_0\right]+2\left(4-6\upxi+\upxi^2\right)\log y_0 +16\left(\upxi-2\right)^2\log\left(\frac{2}{y_0+1}\right)}~.
\end{equation}
As an upper bound, one has the saturation value $-\alpha\ell^2=\omega^2=M^2$, according to Eq.~\eqref{clinL}, which is equivalent to make $\psi=1$~.
Another relevant point regards the brane tension. Higher negative values of $\upxi $ do not bring meaningful influence to the ratio. Thus, one restricts the  analysis dealing with the interval $0\leq \upxi \leq -1$.

\begin{figure}[H]
    \centering
    \includegraphics[width=0.35\textwidth]{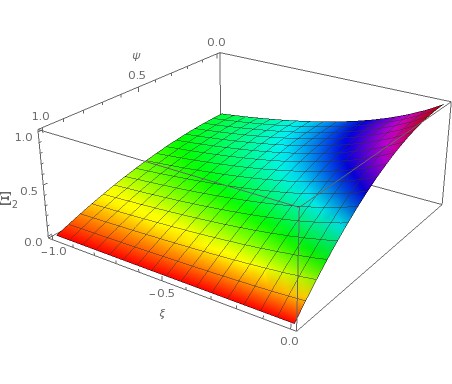}\qquad
    \includegraphics[width=0.35\textwidth]{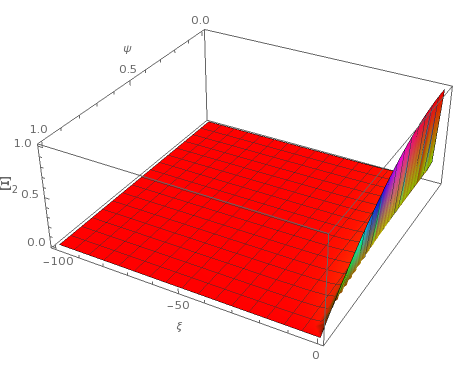}
    \caption{On the left, the behavior of the ratio $\Xi_2$ considering the brane tension varying in the range $-1\leq\upxi\leq 0$. On the right, the low influence of bigger negative values for the brane tension. We adopted $y_0=10^{-10^6}$~.}
    \label{fig:Xi2_psi_xi}
\end{figure}

In Eq.~\eqref{Xi_2}, the ratio $\Xi_2$ is positive, whereas $y_0$ is close to zero. It is very important to emphasize that the limit interferes with the sign of $\Xi_2$. Then, one assumes the lower limit of integration -- a practical small numerical value that can be used -- to investigate the behavior of such ratio. Another interesting point regards the brane tension itself. Bigger negative values of $\upxi$ do not bring meaningful influence to the ratio $\Xi_2$. From now on, the respective analysis of $\Xi_2 $ is restricted to the interval $-1\leq\upxi\leq 0$. Fig.~\ref{fig:Xi2_psi_xi} shows the profile of $\Xi_2$  under the influence of $\upxi$, with $\psi$ assuming different values.

\begin{figure}[H]
    \centering
    \includegraphics[width=0.35\textwidth]{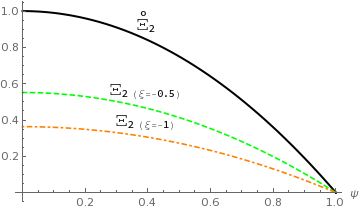}
    \caption{The thick line (the curve in black) shows the ratio of HEE $2^{\textrm{nd}}$-order corrections between a spacetime permeated by an anisotropic fluid and a Schwarzschild one. The dashed line (the curve in green) displays the ratio with the brane tension in an intermediary value, whereas the dot-dashed line goes with the extreme value of the brane tension that has been considered here. In both fixed $\upxi$-curves, one has the anisotropic indicator $\psi$ varying from the null to the saturation value. Again, $y_0=10^{-10^6}$ was adopted.}
    \label{fig:Xi2_psi}
\end{figure}

Fig.~\ref{fig:Xi2_psi} clarifies how the ratio of the $2^{\textrm{nd}}$-order correction is affected by the anisotropic fluid, regarding a fixed brane tension in MGD spacetimes. Of course, with $\upxi=0$ one recovers such correction for a Schwarzschild spacetime, which is also demonstrated.

For smaller values of the anisotropy parameter $\psi$, the ratios $\Xi_2$ are bigger ones. Even though one considers such aspect, the maximum ratio, $\mathring{\Xi}_2=1$, occurs when the brane tension is infinite and $\psi=0$. Within the chosen interval for $\upxi$, the brane under the weakest tension ($\upxi=-1$) leads to a ratio of around $37\%$, since $\psi$ is closest to $0$.
Besides, $\mathcal{S}_2^{\scalebox{.5}{\textrm{Anis}}}$ over $\mathcal{S}_1^{\scalebox{.5}{\textrm{Anis}}}$ is negligible due to stands at $\upepsilon$-order, thus the first law of HEE still holds, since $\delta\mathcal{S}^{\scalebox{.5}{\textrm{Anis}}}=\mathcal{S}^{\scalebox{.5}{\textrm{Anis}}}-\mathcal{S}_0^{\scalebox{.5}{\textrm{Anis}}}\propto M$ even up to the $2^{\textrm{nd}}$-order in the calculations. Holding the first law of HEE is a crucial feature of the HEE paradigm \cite{Sun:2016dch,daRocha:2019pla}. 
This low impact of the anisotropic fluid could be caused by the localization of the boundary. Therefore, the next section treats the opposite situation to localize the boundary, which means settling it very near to the horizon to observe whether the influence this source is meaningful.

\subsubsection{Almost on the horizon}\label{subsect:MGDanisAlmost}

A definitive test must occur while the boundary is localized close the event horizon \cite{daRocha:2019pla,Sun:2016dch} for the background under the influence of an anisotropic fluid, which is characterized by the metric \eqref{metricAnisEspec}. 
\begin{equation}
\mathrm{d}s^2=-\left(\frac{\textrm{r}-\rH}{\textrm{r}}\right)\mathrm{d}t^2 + \left(\frac{\textrm{r}}{\textrm{r}-\rH}\right)\frac{\left(\textrm{r}-\frac{\rH}{2}\right)^2}{\left(\textrm{r}+\rH+\rU\right)\left(\textrm{r}-\rU\right)}\mathrm{d}\textrm{r}^2
+\textrm{r}^2\mathrm{d}\Upomega^2~,\label{mmet}
\end{equation}
where $\rH=2M$ and $\rU=\omega-\rH/2$~. Apparently, there is an extra horizon in  metric (\ref{mmet}). However, sustained by Eq.~\eqref{clinL}, one has no extra singularity when $\rU<0$, which is equivalent to $\omega<M$.

One sets a boundary almost on the horizon adopting a small displacement like $\uprho_0=\upepsilon\sqrt{\rH}$, once $\upepsilon\ll 1$, which models the entangling surface as a circumference, while $\uptheta=\uptheta_0$. In this sense, such setup induces the following metric on the $t$-constant manifold, that is,
\begin{equation}
\mathrm{d}\hat{s}^2=\left[\texttt{p}\texttt{q}\left(
\frac{\mathrm{d}\uprho}{\mathrm{d}\uptheta}\right)^2+\texttt{p}^2\right]\mathrm{d}\uptheta^2
+\texttt{p}^2\sin^2\uptheta\mathrm{d}\varphi^2,
\end{equation}
which considers $\texttt{p}\equiv\texttt{p}(\uprho)=\uprho^2+\rH$ and
$\texttt{q}\equiv\texttt{q}(\uprho)=\frac{4\left(\uprho^2+\frac{3}{2}\rH\right)^2}{\left(\uprho^2+2\rH+\rU\right)\left(\uprho^2+\rH+\rU\right)}$, while $\uprho\equiv\uprho(\uptheta)$.

To determine the HEE, at first, it is necessary to compute $\uprho$, which leads to the minimization of the surface area, \emph{i.e.},
\begin{equation}\label{area-gammaA}
\widetilde{\text{Area}}(\gamma_A)= 2\pi\int_{y_0}^{1} \mathrm{d}y~\tilde{\mathcal{I}},  
\end{equation}
where $\tilde{\mathcal{I}}=\left[\texttt{p}\texttt{q}(1-y^2)
\dot{\uprho}^2+\texttt{p}^2\right]^{1/2}$. Once again, the change of variable  $y=\cos\uptheta$, has been employed to get $\uprho\equiv\uprho(y)$. Therefore, the minimization of Eq.~\eqref{area-gammaA} with respect to $y$ gives
\begin{equation}\label{ODE-aniso-almost}
    (1-y^2)\left[2y\texttt{q}^2\dot{\uprho}^3+\left(5\texttt{q}\frac{\mathrm{d}\texttt{p}}{\mathrm{d}\uprho}-\texttt{p}\frac{\mathrm{d}\texttt{q}}{\mathrm{d}\uprho}\right)\dot{\uprho}^2\right] + 2\texttt{p}\left[2\frac{\mathrm{d}\texttt{p}}{\mathrm{d}\uprho}+2y\texttt{q}\dot{\uprho}-2\texttt{q}\left(1-y^2\right)\ddot{\uprho}\right]=0~.
\end{equation}

The endeavor of looking for an analytical solution of Eq. \eqref{ODE-aniso-almost} drops off, in front of the strong nonlinearity of such differential equation. Hence, one must implement a perturbative method, adopting the expansion
\begin{equation}\label{rho-almost}
    \uprho(y)=\upepsilon\uprho_1(y)+\upepsilon^2\uprho_2(y)~,
\end{equation}
since $\uprho_1(y_0)=\sqrt{\rH}$ and $\uprho_2(y_0)=0$, considering the boundary condition $\uprho(y_0)=0$.

The $0^{\rm th}$-order term in Eq.~\eqref{rho-almost} is absent just to avoid an area having a value greater than one valued at the point $(\uprho=\uprho_0,\uptheta=\uptheta_0)$. 
 Thus, searching for the $\uprho$-functions up to $2^{\textrm{nd}}$-order, one inserts Eq.~\eqref{rho-almost} into Eq.~\eqref{ODE-aniso-almost}. Thus, at $1^{\rm st}$-order in $\upepsilon$, one gets
\begin{equation} \label{ODE1Aniso-almost}
(y^2-1)\ddot{\uprho}_1+2y\dot{\uprho}_1+\frac{1}{\kappa}\uprho_1=0~,
\end{equation}
where
\begin{equation}\label{kappa-inv}
\frac{1}{\kappa}=\frac{4}{9}\left(1-\frac{\rU}{\rH}\right)\left(2+\frac{\rU}{\rH}\right)~.
\end{equation}
In this manner, one can notice the influence of the main parameters of the gravitational decoupling within the minimal geometric deformation under an anisotropic fluid as a source.

One determines the solution for Eq.~\eqref{ODE1Aniso-almost}, which is
\begin{equation}
   \uprho_1(y)=\frac{\sqrt{\rH}}{\mathrm{P}_\uptau(y_0)}\mathrm{P}_\uptau(y) 
\end{equation}
with $\uptau=\frac12\left(-1+\sqrt{1-\frac{4}{\kappa})}\right)$ and $\mathrm{P}_\uptau(y)$ denotes the Legendre polynomial of $1^{\textrm{st}}$ kind. This solution is regular at $y=1$, with boundary condition $\uprho_1(y_0)=\sqrt{\rH}$~.

Moving to the next order, that is, towards the $2^{\rm nd}$-order in $\upepsilon$, one arrives at a Legendre equation that is similar to Eq.~\eqref{ODE1Aniso-almost}, reading 
\begin{equation}\label{ODE2Aniso-almost}
(y^2-1)\ddot{\uprho}_2+2y\dot{\uprho}_2+\frac{1}{\kappa}\uprho_2=0~,
\end{equation}
which has a solution like $\uprho_2(y)=\mathcal{B}\mathrm{P}_\uptau(y)$. By the other side, a boundary condition $\uprho_2(y_0)=0$ implies $\mathcal{B}=0$. Hence, $\uprho_2(y)=0$, which simplifies the $\uprho$-function uniquely to the $1^{\rm st}$-order in $\upepsilon$.

\begin{figure}[H]
    \centering
    \includegraphics[scale=.5]{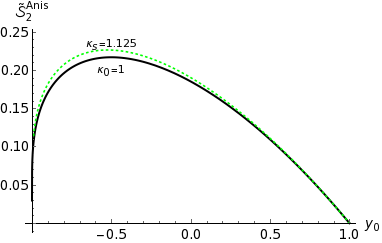}
    \caption{The evolution of the HEE $2^{\rm nd}$-order correction, in units of $\pi{\rH}\uprho_0^2$, related to two values of the anisotropic parameter, $\kappa_\textrm{s}$ -- the dotted green curve -- and $\kappa_0$ -- the thick black curve, according to the size of the subsystem $y_0$.}
    \label{fig:S2_kMaxAndMin_y0_aniso_almost}
\end{figure}

As a final development, with the $\uprho$-function in hands, one calculates the area of the entangling surface. Firstly, expanding the integrand of Eq.~\eqref{area-gammaA} in $\upepsilon$, which yields 
\begin{equation}\label{I-expanded}
\tilde{\mathcal{I}}=2\pi\rH^2 +4\pi \rH\left[(1-y^2)\dot{\uprho}_1^2+\uprho_1^2\right]\upepsilon^2+\ldots.
\end{equation}
Secondly, one substitutes the Eq.~\eqref{I-expanded} into Eq.~\eqref{area-gammaA}, executes the expansion of $\widetilde{\rm Area}$, which is $
\widetilde{\rm Area}=\widetilde{\rm Area}_0+\widetilde{\rm Area}_1+\widetilde{\rm Area}_2+\cdots~$~. Restricting to the $2^{\textrm{nd}}$-order, the correspondent HEE has the following contributions:
\begin{align}
\tilde{\mathcal{S}}_0^{\scalebox{.5}{Anis}} &= \frac{\pi\rH^2}{2}\left(1-y_0\right),\nonumber\\
\tilde{\mathcal{S}}_1^{\scalebox{.5}{Anis}}&= 0,\nonumber\\
\label{S2Aniso_almost}
\tilde{\mathcal{S}}_2^{\scalebox{.5}{Anis}}&=\frac{\pi\rH\uprho_0^2}{\mathrm{P}_\uptau^2(y_0)}
\int_{y_0}^1 \mathrm{d}y\,\left[\kappa\left(1-y^2\right)\dot{\mathrm{P}}_\uptau^2(y)+\mathrm{P}_\uptau^2(y)\right]. 
\end{align}
Without an efficient method to calculate analytical solutions of $\tilde{\mathcal{S}}_2^{\scalebox{.5}{Anis}}$, a clever procedure to deal with it consists of using numerical computation techniques. Indeed, numerically, Fig.~\ref{fig:S2_kMaxAndMin_y0_aniso_almost} shows two specific values of $\kappa$. In fact, the saturation value $(\kappa_\textrm{s}=1.125)$ and the value $\kappa_0=1 $ that recovers the Schwarzschild solution.

\begin{figure}[H]
    \centering
    \includegraphics[scale=.45]{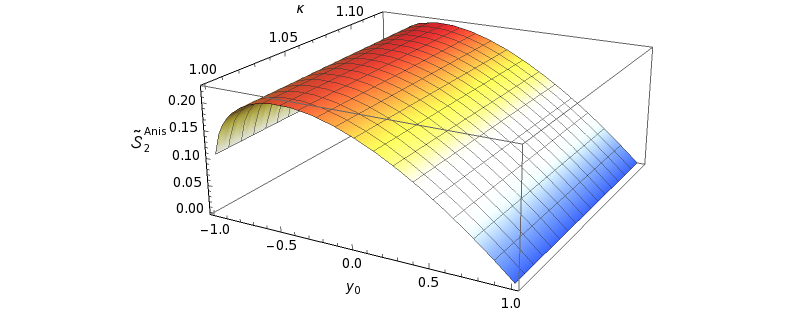}
    \caption{The evolution of the $2^{\rm nd}$-order correction of HEE, in units of $\pi\mathring{r}\uprho_0^2$, related to the $\kappa$ and $y_0$ considering the total range to each one of them.}
    \label{fig:S2_k_y0_aniso_almost_temp}
\end{figure}

A general scenario based on this order of correction is displayed in Fig.~\ref{fig:S2_k_y0_aniso_almost_temp}, considering a valid range for $\kappa$ and the complete size of the subsystem $y_0$.
One can realize that the $\tilde{\mathcal{S}}_2^{\textrm{Anis}}$ maximum value occurs when one takes into account the lower limit of integration at $y_0\approx -0.53$ and the saturation value $\kappa_\textrm{s}=1.125$. The shape of the 3D surface showed in Fig.~\ref{fig:S2_k_y0_aniso_almost_temp} has a similar behavior, for all $\kappa$ inside the range of validity.

\begin{figure}[H]
    \centering
    \includegraphics[scale=.35]{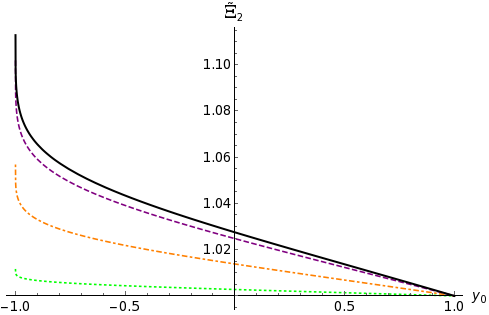}
    \caption{The evolution of the ratio considering HEE $2^{\rm nd}$-order corrections, in units of $\pi\mathring{r}\uprho_0^2$, considering $10\%$ (the dotted green curve), $50\%$ (the dot-dashed orange curve), $90\%$ (the dashed purple curve) and $100\%$ (the black thick curve) of attenuation of the saturation value, \emph{i.e.}, $\zeta_\textrm{s}=-1/9$ (or $\kappa_\textrm{s}=1.125$).}
    \label{fig:Xi2Tilde_aniso_almost}
\end{figure}

Regarding  the ratio of $2^{\textrm{nd}}$-order corrections of HEE, taking the entangling surface almost on the horizon,  Fig.~\ref{fig:Xi2Tilde_aniso_almost} shows the ratio $\tilde{\Xi}_2=\tilde{\mathcal{S}}_2^{\scalebox{0.5}{\textrm{Anis Sat}}}/\tilde{\mathcal{S}}_2^{\scalebox{.5}{MGD}}$, where $\tilde{\mathcal{S}}_2^{\scalebox{.5}{MGD}}$ is the $2^{\textrm{nd}}$-order correction of HEE for a MGD spacetime\footnote{
From \cite{daRocha:2019pla}, one recalls that
\begin{equation}
\tilde{\mathcal{S}}_2^{\scalebox{.5}{MGD}}=\frac{\pi\mathring{r}\uprho_0^2}{\mathrm{P}_\eta^2(y_0)}\int_{y_0}^1 \mathrm{d}y\,\left[\left(\frac{1-y^2}{1+\zeta}\right)\dot{\mathrm{P}}_\eta^2(y)+\mathrm{P}_\eta^2(y)\right]\nonumber~,
\end{equation}
where $\mathring{r}=2M$ and $\eta=1/2\left(-1+\sqrt{-3-4\zeta}\right)$.}, established by a calibrated brane tension (represented by particular $\zeta$ values\footnote{It has been used $\zeta$ in this present work instead $\alpha$, which was used in \cite{daRocha:2019pla}, to avoid confusion with the $\upalpha$ parameter, employed in  Sect.~\ref{sect:MGD-setup}.}), whereas $\tilde{\mathcal{S}}_2^{\scalebox{0.5}{\textrm{Anis~Sat}}}$ stands for the saturated $2^{\textrm{nd}}$-order correction of HEE for the spacetime permeated by an anisotropic fluid. Besides, taking the saturation value $\kappa_\textrm{s}=1.125$, one gets $\zeta_\textrm{s}=-1/9\approx~-0.1111$, which means that one has a saturated correspondence between a background with an extra anisotropic fluid and a generic MGD spacetime. Each curve adopts a percentual attenuation of $\zeta_\textrm{s}$.

One might see clearly the influence of the anisotropy while one looks for the saturation value $\kappa_\textrm{s}=1.125$ and comparing with the Schwarzschild case. Another important feature shown in Fig.~\ref{fig:Xi2Tilde_aniso_almost} is that higher ratio occurs with the totality of the interval of integration instead of using the limit $y_0\approx-0.53$, where the maximum $2^\textrm{nd}$-order contribution happens for the saturation value $\kappa_\textrm{s}=1.125$.

\begin{figure}[H]
    \centering
    \includegraphics[scale=.45]{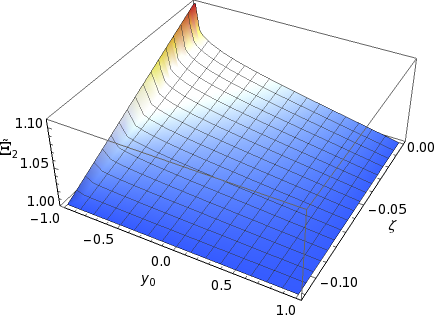}
    \caption{The general evolution of the ratio considering $2^{\rm nd}$-order corrections of HEE, in units of $\pi\mathring{r}\uprho_0^2$, considering $0\leqslant\zeta\leqslant -1/9$ and the size of the subsystem.}
    \label{fig:Xi2Tilde_aniso_almost_general}
\end{figure}
Finally, Fig.~\ref{fig:Xi2Tilde_aniso_almost_general} displays the complete spectrum of $\zeta$ values, regarding all sizes of the subsystem to widely clarify  the behavior of the ratio $\tilde{\Xi}_2$~. Clearly, one notices that the most accentuated growing of the ratio $\tilde{\Xi}_2$ occurs  in the sector with the largest size of the subsystem and values of  $\zeta$ close to zero, which is profoundly related to high values of the brane tension.

\section{Final Remarks and conclusions}\label{sect:Final}

The MGD procedure has been applied to derive new solutions, acting as a solid and reliable mechanism to calculate high energy corrections to the GR. Therefore, the addition of an extra anisotropic source into the stress-energy tensor can originate  a family of anisotropic black holes. This important line is combined with another one, which is delineated by the HEE conjecture for asymptotically flat spacetimes. Hence, the HEE was calculated for an anisotropic black hole considering boundaries -- the situs to settle the entangling surfaces -- far from the horizon and almost on it, 
permitting to define corresponding dual field theories. In the absence of a complete analytical solution, an expansion was used to determine the HEE up to the $2^\textrm{nd}$-order in both scenarios that those boundaries were settled up. In addition, comparative investigations were implemented, considering order-by-order  corrections of the HEE for MGD and Schwarzschild spacetimes with respect to  anisotropic black hole.

For a boundary localized far from the horizon, the HEE of an anisotropic black hole is different of the HEE for an original MGD black hole, after one looks their respective corrections beyond the $0^\textrm{th}$-order. The $1^\textrm{st}$-order of correction does not have any relationship with the MGD parameter driven by the brane tension,  shown by Eq.~\eqref{S1anisoBeyond}. By the way, it is the same value for the same order of correction to the HEE for a Schwarzschild black hole. The parameter $\upepsilon$ takes a large radial distance to set the boundary, then can be the reason for such equality. Even though, comparing the $1^\textrm{st}$-order corrections of HEE between anisotropic black holes and MGD ones, there is an important contribution caused by the brane tension carried by the parameter $\upxi$ in Eq.~\eqref{S1anisoBeyond}, and has featured in the ratio $\Xi_1$ with its sole dependence on such parameter. Thus, smaller brane tension values rise the impact of an anisotropic source, in front of a standard MGD formulation, clearly demonstrated by Eq.~\eqref{Xi_1}.

Towards the next order, a wider scenario rises. In fact, two parameters enter in the respective ratio $\Xi_2$, displayed in Eq.~\eqref{Xi_2}. First, the full scenario about the influence of the brane tension in the anisotropy parameter was studied, restricting to the interval $-1\leqslant\upxi\leqslant 0$, where it is possible to see the major contribution of the brane tension aggregated with the anisotropy parameter $\psi$.  Fig.~\ref{fig:Xi2_psi} illustrates the influence of $\psi$ and $\upxi$. Immediately, it is possible to notice that the brane tension is rather than the weighting of the source with an anisotropic fluid, while $\psi$ is small. On the other hand, very close to the saturation value of $\psi$, all ratios go to $0$, which implies the high weight of the anisotropic fluid in the attenuation of the $2^\textrm{nd}$-order correction of HEE. Besides, considering the values at the extrema of the $\upxi$-interval, the ratio was obtained, indicating that the lower the brane tension, the bigger the ratio is. The choice of the largest distance to settle the entangling surface was imperative to determine $2^\textrm{nd}$-order correction of HEE for the anisotropic black hole, even less than the correspondent order of correction of HEE for pure MGD spacetimes under low brane tension values and also for the Schwarzschild black hole, prompting the calculations of HEE closer to the event horizon. At the end, the first law of HEE holds -- since $\delta\mathcal{S}^\textrm{Anis}\propto M$ -- which is a crucial point based on the HEE paradigm.

Almost on the horizon, the boundary there establishes a HEE up to $2^\textrm{nd}$-order correction in an adapted $\upepsilon$-expansion, employed to compute the minimal area of a respective entangling surface. The $0^\textrm{th}$-order of HEE matches with the same order of correction of the HEE for the MGD spacetime as well as for the Schwarzschild one. The vanishing of the $1^\textrm{st}$-order is also determined in the present case. The $2^\textrm{nd}$-order correction of HEE has been computed numerically, considering the saturation value $\kappa_\textrm{s}=1.125$ that corresponds to the upper bound of $-\alpha\ell^2\leqslant M^2$, which has been imposed by the requirement of no extra singularity. It heads to an important restriction, when comparing with the MGD spacetime. In fact, additional anisotropic sources bring a restrictive range for brane tensions, that is, the saturation value $\kappa_\textsc{s}=1.125$ gives a lower bound $\zeta_\textrm{s}=-1/9$ working as a limit for the brane tension. Obeying this physical constraint, the curve of values for such order of correction is displaced when compared with that correspondent order of correction of HEE for a Schwarzschild spacetime, as shown in Fig.~\ref{fig:S2_kMaxAndMin_y0_aniso_almost} and Fig.~\ref{fig:S2_k_y0_aniso_almost_temp}. This last one emphasizes the influence of the size of the subsystem with the lower limit $y_0$. Concerning the maximum value of the $2^\textrm{nd}$-order correction, it has occurred with an inferior limit at $y_0\approx-0.53$ (while $y=1$ works as the superior one) to the size of the subsystem -- see Fig.~\ref{fig:S2_kMaxAndMin_y0_aniso_almost}.

Adding on, the ratio $\tilde{\Xi}_2$ is plotted in Fig.~\ref{fig:Xi2Tilde_aniso_almost} to demonstrate the impact of the anisotropy parameter on the $2^\textrm{nd}$-order correction of HEE,  when compared to the same order of correction of HEE for both the  Schwarzschild and the MGD solutions. The full size of the dual subsystem leads to the ratio $\tilde{\Xi}_2$ as higher as the saturation value gets attenuation. If one looks the $2^\textrm{nd}$-order corrections of HEE for a saturated anisotropic black hole and the Schwarzschild one, there is an $11\%$ increment of the ratio $\tilde{\Xi}_2$ with the largest possible size of the subsystem, against a $4.4\%$ increment with the lower  extremum at $y_0=-0.53$, which means a $247\%$ increment from the minor percentage. Besides, Fig.~\ref{fig:Xi2Tilde_aniso_almost_general} reveals the accentuated growing of $\tilde{\Xi}_2$ after tackling the largest size of the subsystem and lower values of the anisotropy parameter $\kappa$, which is linked with the brane tension parameter $\zeta$\footnote{To make sense, a correspondence with the corresponding brane tension parameter $\zeta$ from the MGD analysis was implemented, yielding $\kappa=1/(1+\zeta)$. In this manner, the anisotropy parameter $\kappa$ is formally related to the brane tension parameter $\zeta$.}.

\paragraph*{Acknowledgments:}\;    RdR~is grateful to FAPESP (Grant No.  2017/18897-8) and to the National Council for Scientific and Technological Development  -- CNPq (Grants No. 406134/2018-9, No. 303390/2019-0 and No.  303293/2015-2), for partial financial support. AAT thanks to PNPD -- CAPES -- UFABC (Proc. No. 88887.338076/2019-00) and PNPD -- CAPES -- UFF (Proc. No. 88887.473671/2020-00).

\bibliographystyle{iopart-num}
\bibliography{1bib_HEE}

\end{document}